\documentclass[twocolumn,amsmath,amssymb,10pt,prl]{revtex4}

\usepackage{graphicx}
\usepackage{bm}
\usepackage{color}
\usepackage{ulem} 

\newcommand{\tcr}[1]{\textcolor{black}{#1}}

\begin{document}
\title{Entangling Macroscopic Light States by Delocalized Photon Addition}

\author{Nicola Biagi, Luca S. Costanzo, Marco Bellini$^{*}$, and Alessandro Zavatta†}

\affiliation{Istituto Nazionale di Ottica (CNR-INO), L.go E. Fermi 6, 50125 Florence, Italy\\
LENS and Department of Physics $\&$ Astronomy, University of Firenze, 50019 Sesto Fiorentino, Florence, Italy}

\bigskip
\date{\today}

\begin{abstract}
\tcr{We present a scheme, based on the delocalized heralded addition of a single photon, to entangle two or more distinct field modes, each containing arbitrary light states. A high degree of entanglement can in principle endure light states of macroscopic intensities and is expected to be particularly robust against losses. We experimentally establish and measure significant entanglement between two identical weak laser pulses containing up to 60 photons each.}
\end{abstract}

\maketitle
Entanglement is a distinctive feature of quantum mechanics marking the most striking deviations of its predictions from those of classical physics. Although it has been widely experimentally demonstrated in several microscopic systems, with recent achievements including the loophole-free violation of Bell's inequalities \cite{bell1,bell2}, generating and detecting entanglement between larger and larger objects is an increasingly difficult task, whose experimental limits are highly worth investigating. 

Recent optical experiments have demonstrated the generation of a so-called 'micro-macro' entanglement \cite{bruno13,lvovsky13}, where one part of a system in a microscopic superposition of vacuum and one-photon states is entangled with another part containing a macroscopic mean number of photons. Here we propose and test a different scheme, which allows one to entangle two or more modes, each containing arbitrarily large numbers of photons. Being this 'macro-macro' entanglement independent of the size of the entangled partners and surprisingly robust against losses~\cite{macrogisin}, it is therefore an exceptional test-bed for studying the resilience and detectability of entanglement for states of growing macroscopicity and as a potential means of transmission over long distances.

Our basic ingredient is the possibility of performing the coherent delocalized addition of a single photon over different modes. Photon addition (by the creation operator $\hat{a}^{\dagger}$) and subtraction (by the annihilation operator $\hat{a}$) have already demonstrated to be extremely useful for performing operations normally unavailable in the realm of Gaussian quantum optics \cite{kerr,noiseless,progopt}. Photon subtraction from a single-mode photonic state can de-Gaussify it~\cite{degauss} and enhance its nonclassicality~\cite{catsub}. Moreover, it can increase and distill existing two-mode entanglement \cite{entinc,entdist}. On the other hand, photon addition has the unique capability of creating (single-mode) nonclassicality~\cite{sciadd,qpn} and (multimode) entanglement, whatever the input states.
\begin{figure}[h]
	\begin{center}
		\includegraphics[width=0.75\linewidth,angle=0]{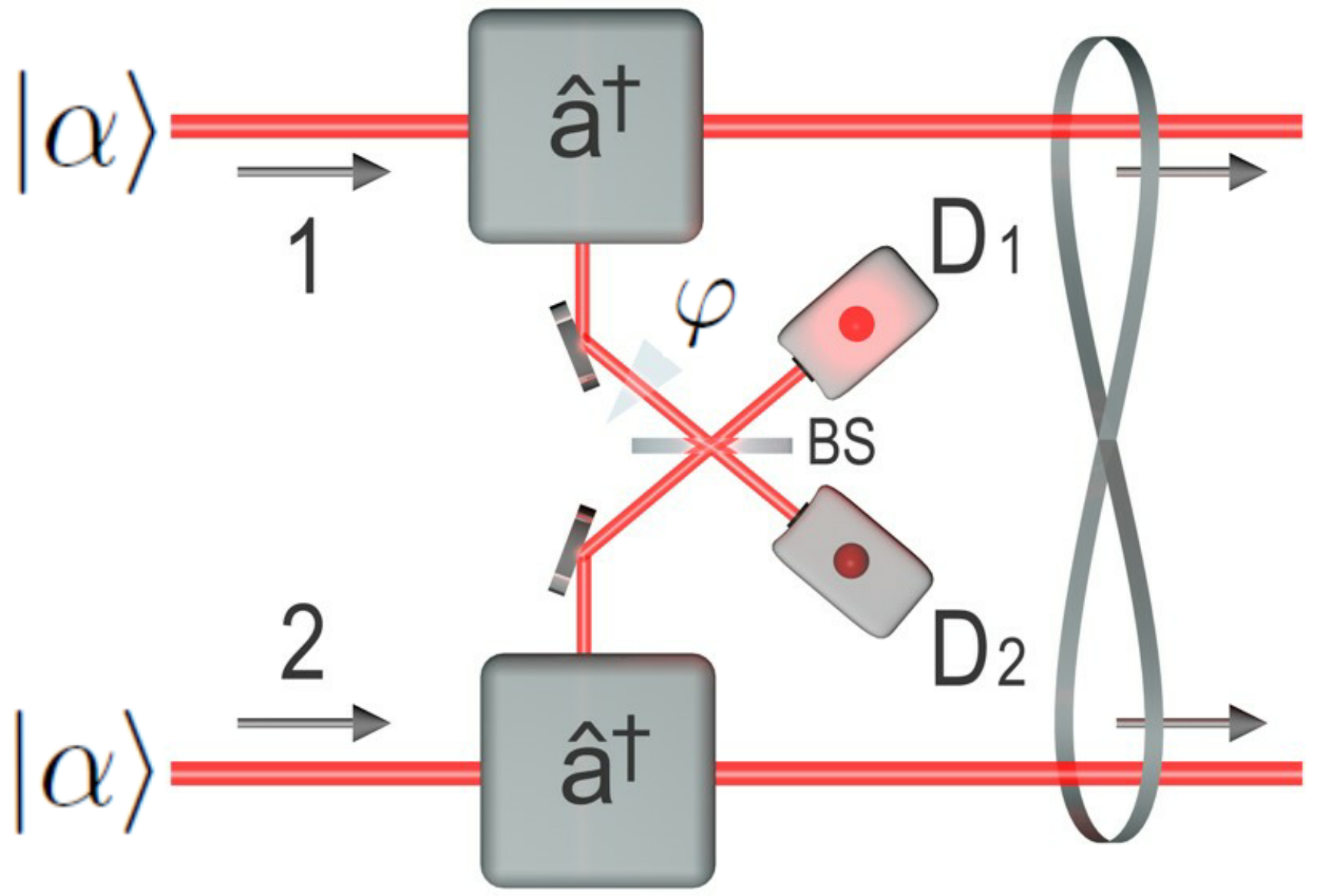}
		\caption{Conceptual experimental scheme to perform a coherent single-photon addition on two different input modes, both containing a coherent state $\vert \alpha\rangle$. A click in a single-photon detector $D_1$, placed after a balanced beam splitter BS mixing the herald modes of two photon-addition modules based on parametric down-conversion (PDC) \cite{sciadd,progopt}, generates entanglement between the two output modes.}
		\label{fig:expscheme}
	\end{center}
\end{figure}

In particular, the coherent addition of a single photon to two distinct field modes, $1$ and $2$, entangles them, and the entanglement produced by a balanced superposition of the kind $\hat{a}^{\dagger}_1+ e^{i\varphi}\hat{a}^{\dagger}_2$ depends on the states of light already present in the two modes before the operation. If both are originally in a vacuum state, one simply obtains a single-photon mode-entangled state \cite{remoteprep} of the kind $\vert \psi\rangle_{12} =  ( \vert 1\rangle_1 \vert 0\rangle_2 + e^{i\varphi}\vert 0\rangle_1 \vert 1\rangle_2)/\sqrt{2}$. If different quantum states originally populate the field modes, the state resulting from delocalized photon addition may present different features. For example, injecting a vacuum and a coherent state in the two input modes gives rise to a so-called hybrid discrete/continuous-variable entanglement of the two output modes \cite{hybrident}.

Here we study the effect of delocalized single-photon addition on two input modes containing identical coherent states $\vert \alpha\rangle$, as schematically illustrated in Fig.\ref{fig:expscheme}. 
The general entangled state produced by this operation can be written as follows:
\begin{eqnarray}
\vert \psi_{\varphi}(\alpha)\rangle_{12}
&=& \Big(\hat{a}^{\dagger}_1 \vert \alpha\rangle_1 \vert \alpha\rangle_2+ e^{i \varphi}\vert \alpha\rangle_1 \hat{a}^{\dagger}_2\vert \alpha\rangle_2\Big)/\sqrt{\mathcal{N}}\nonumber\\
& =&\Big[\hat D_1(\alpha)\hat D_2(\alpha) \Big( \vert 1\rangle_1 \vert 0\rangle_2 + e^{i \varphi}\vert 0\rangle_1 \vert 1\rangle_2\Big) 
\nonumber\\
&+&\alpha^*(1+e^{i \varphi})\vert \alpha\rangle_1 \vert\alpha\rangle_2\Big]/\sqrt{\mathcal{N}}
\label{eq:displace}
\end{eqnarray}
with the normalization factor $\mathcal{N} = 2[1+|\alpha|^2(1+\cos\varphi)]$ and the phase-space displacement operator $\hat D(\alpha)=e^{\alpha \hat{a}^{\dagger}-\alpha^* \hat{a}}$. Already in this simple case of balanced photon addition, the output state shows a very rich structure resulting from the coherent contribution of an entangled and a separable part with adjustable weights depending on the superposition phase $\varphi$ and on the amplitude $\alpha$ of the coherent states. One can quantify the degree of entanglement in a state described by the density matrix $\hat{\rho}$ by calculating the so-called negativity of the partial transpose (NPT) \cite{npt}, which is proportional to the sum of the negative eigenvalues $\lambda_i^-$ of the partially transposed density matrix $\hat{\rho}^{PT}$, and is therefore defined as:
\begin{equation}
NPT(\hat{\rho})=-2\sum_i \lambda_i^- ,
\label{eq:NPTdef}
\end{equation} 
where the factor 2 guarantees that $0\leq NPT(\hat{\rho}) \leq 1$. For the state described by Eq.(\ref{eq:displace}), the NPT is calculated to be:
\begin{equation}
NPT(\vert \psi_{\varphi}(\alpha)\rangle\langle \psi_{\varphi}(\alpha)\vert) = \frac{1}{1+ \vert\alpha\vert ^2(1+\cos \varphi)}.
\label{eq:NPTteo}
\end{equation} 

It is easy to see that, in the extreme case of an even superposition with $\varphi=0$, the degree of entanglement of the state quickly deteriorates for increasing $\alpha$, due to the large contribution of the separable fraction in Eq.(\ref{eq:displace}).
\begin{figure}[h]
	\includegraphics[width=0.9\linewidth,angle=0]{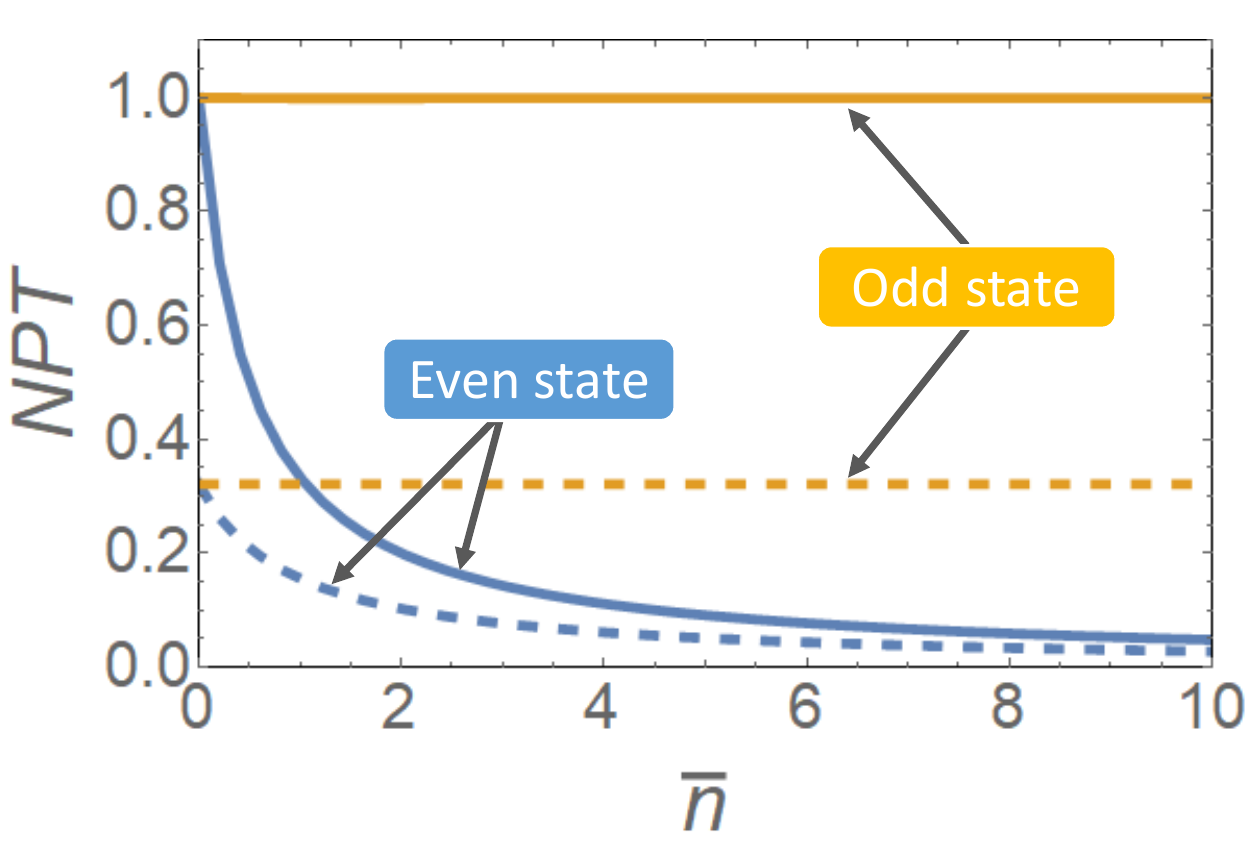}
	\caption{Theoretical entanglement (quantified via the negativity of the partial transpose, NPT) of the odd and even entangled states (yellow and blue solid curves, respectively), according to Eq.(\ref{eq:NPTteo}). Dashed curves present the numerically calculated NPT behavior when both the modes are subjected to 40\% of losses.} \label{fig:negat}
\end{figure}
However, the entangled contribution can be continuously tuned by varying the superposition phase $\varphi$, until the other extreme condition of $\varphi=\pi$ is reached. In this case, the odd superposition entangled state
\begin{equation}
\vert \psi_{\pi}(\alpha)\rangle_{12} = \frac{1}{\sqrt{2}} (\hat{a}^{\dagger}_1 \vert \alpha\rangle_1 \vert \alpha\rangle_2 -\vert \alpha\rangle_1 \hat{a}^{\dagger}_2\vert \alpha\rangle_2),
\label{eq:oddstate}
\end{equation}
is seen to be equivalent to the result of an equal phase-space displacement operation $\hat D(\alpha)$ on both modes of a single-photon mode-entangled state. As such, it is expected to maintain constant entanglement independently of the amplitude of the input coherent states (see Fig.\ref{fig:negat}). Ideally, a high degree of entanglement (NPT=1) should thus be observable in the $\vert \psi_{\pi}(\alpha)\rangle$ state even between two modes initially containing large, and possibly macroscopic, mean photon numbers $\bar n =|\alpha|^2$. More interestingly, this behavior is preserved even when the states are affected by an overall limited efficiency $\eta$, accounting for channel transmission losses and detection inefficiency. This is shown by the dashed curves of Fig.~\ref{fig:negat}, where the maximum entanglement is given by:
\begin{equation}
NPT_{max}(\eta)= \frac{1}{\eta -1 +\sqrt{(\eta -1)^2+\eta^2 }}.
\end{equation}

The odd superposition entangled states (\ref{eq:oddstate}) have already been theoretically discussed in Ref.\cite{macrogisin}, and the 'micro-macro' entanglement experiments of~\cite{bruno13,lvovsky13} are scaled-down realizations of such a proposal, where displacement is performed on just one of the two modes. Moreover, in those experiments entanglement is finally verified by 'un-doing' the macroscopicity and optically displacing the state back to the ($\vert 0\rangle$, $\vert 1\rangle$) Fock subspace before measurements, as recently done also in Ref.\cite{sychev18}. Differently from the approaches of Refs.\cite{bruno13,lvovsky13}, here we generate real 'macro-macro' entanglement, where both modes contain states of macroscopic size that are also macroscopically distinguishable from each other. Furthermore, and differently also from Ref.\cite{sychev18}, here we perform two-mode homodyne detection of the entire macroscopic states, in principle allowing us to directly evaluate their entanglement for arbitrary sizes.
\begin{figure}[h!]
	\begin{center}
		\includegraphics[width=.86\linewidth]{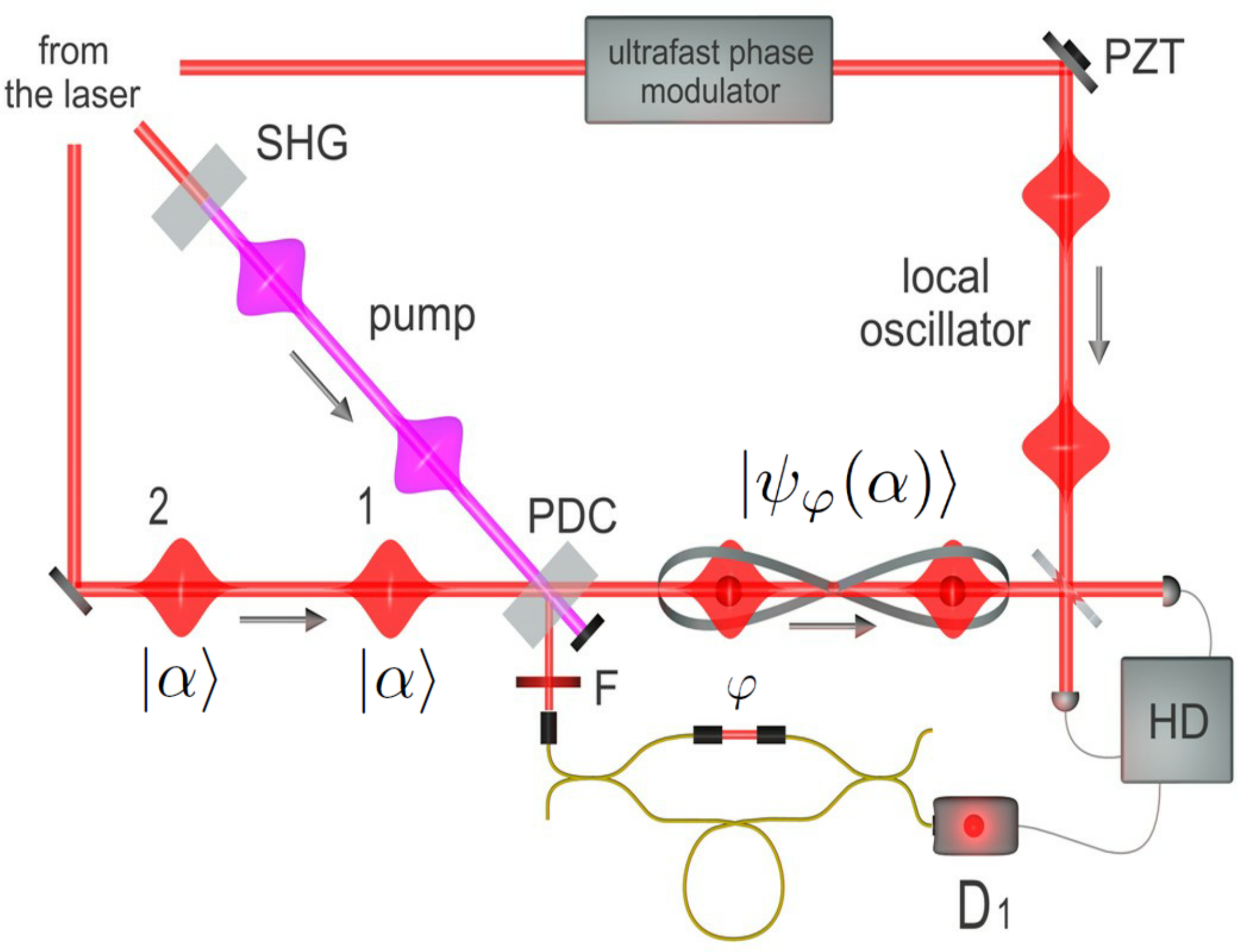}
		\caption{Experimental setup for coherent single-photon addition on two different input temporal modes, both containing a coherent state $\vert \alpha\rangle$. It is based on a mode-locked laser emitting 1.5-ps pulses at 786 nm with a repetition rate of 81 MHz, providing: the local oscillator (LO) for homodyne detection, the pump for a parametric down-conversion (PDC) process after a frequency-doubling stage (SHG), and the seed coherent states for photon addition in the PDC crystal. PDC photons emitted in the idler channel pass through a set of spectral and spatial filters (F) before entering an unbalanced, fiber-based, Mach-Zehnder interferometer. A detection event by the single-photon detector ($D_1$) placed at one of the interferometer outputs heralds the successful implementation of a delocalized single-photon addition, meaning that entanglement has been conditionally generated between the two temporal modes. The state superposition phase $\varphi$ is remotely controlled by varying the relative phase between the interferometer arms via a fine adjustment of an air-gap length. A feedback loop based on the interference of a counterpropagating pulse train injected in the unused interferometer output port provides phase stabilization.}
		\label{fig:trueexp}
	\end{center}
\end{figure}

We use the temporal-mode version (see Fig.\ref{fig:trueexp}) of the setup illustrated in Fig.\ref{fig:expscheme} for generating states of the form of Eq.(\ref{eq:oddstate}) experimentally. In this case, the two devices for single-photon addition are replaced by a single one, operating on two different traveling temporal modes. The coherent superposition of photon additions on mode $1$ or $2$ is obtained by allowing the herald photon from the addition device (based on stimulated PDC \cite{progopt}) to travel two indistinguishable paths of different length towards the herald detector \cite{remoteprep}.
In principle, besides its higher phase stability, the temporal-mode version of the experiment has the fundamental additional advantage of an easy scalability, because it allows one to increase the number of involved modes without a corresponding multiplication of photon addition devices and detectors.

We inject several coherent state amplitudes $\alpha$ at the input, and perform a quantum tomographic analysis of the final output states based on two-mode time-domain homodyne detection~\cite{Zavatta02}. Here, differently from \cite{remoteprep, hybrident}, the phases of the two local oscillator pulses are independently changed in the $[0,\pi]$ interval by controlling their global phase via a piezo-mounted mirror (PZT), and their relative phase by means of a fast electro-optic modulator. An ultraprecise timing system, based on a digital synthesizer, is used to generate the modulator driving signal, which is locked to the laser pulses, and to synchronize the acquisition system. The reconstructed two-mode density matrices are then used to calculate their NPT and extract the degree of entanglement of the states as a function of their macroscopicity.

At this point, it is worth noting that a faithful representation of states like those of Eq.~(\ref{eq:oddstate}) requires the reconstruction of a number of density matrix elements that grows extremely fast with the coherent state amplitude $|\alpha|$. For example, already for $|\alpha|\approx 7$, corresponding to a mean photon number of $\bar n=|\alpha|^2\approx 50$ photons per mode, at least $3\times 10^7$ density matrix elements need to be calculated. Since the brute force approach of full density matrix reconstruction has no hope to succeed with such a huge number of elements, we adopt two different strategies to restrict the reconstructed subspace.

\begin{figure}[h]
	\begin{center}
		\includegraphics[width=0.8\linewidth]{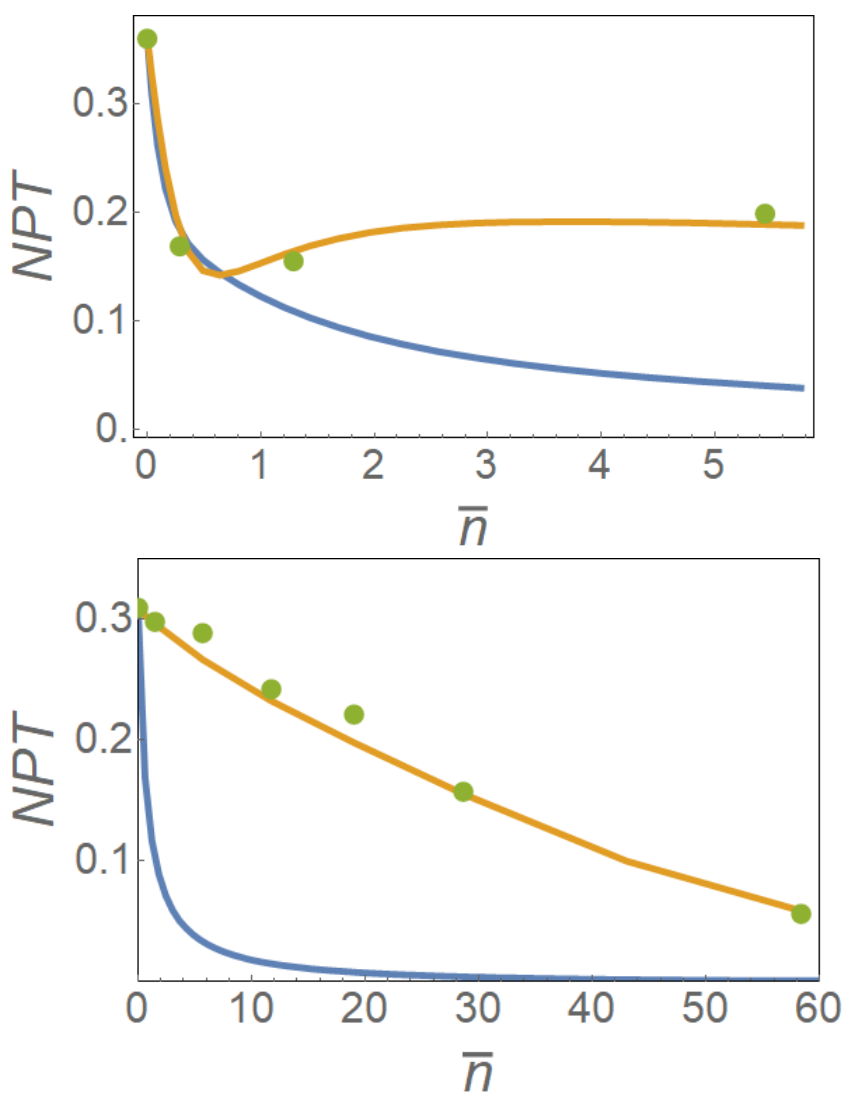}
		\caption{Experimental (green dots) and calculated NPT for the odd entangled states (yellow solid curves) as a function of the mean photon number $\bar n$ (the calculated NPT curves for even entangled states are also shown in blue for reference). Upper panel: ``global-phase-averaged tomography''. The theoretical curves are calculated with a detection efficiency $\eta = 68\%$. Bottom panel: Tomography based on correlated quadrature fluctuations shows significant entanglement between modes with up to about 60 photons each. The theoretical curves are calculated by considering a detection efficiency $\eta = 64\%$ together with a noise of $\pi/100$ on both the state phase $\varphi$ and the LO global phase. In addition, the observed value of entanglement is mainly limited by the state preparation efficiency, degrading with increasing mean photon number due to a nonperfect Mach-Zehnder visibility of 99.6\%. Statistical errors are evaluated with a bootstrap method using 100 resampled quadrature data sets. For each data set we reconstructed the density matrix and calculated the NPT. Error bars in the plot are estimated as the standard deviation of these NPT values and are smaller than the corresponding dot symbols.}
		\label{fig:expnegat}
	\end{center}
\end{figure}

In the first method, the global LO phase for homodyne detection is actively randomized while the relative phase between the two temporal LO modes is scanned in a controlled way over 9 different values. About 50 000 quadrature measurements per mode are performed for each value of the relative LO phase, and only the density matrix elements diagonal with respect to the LO global phase are then reconstructed by means of an iterative maximum likelihood algorithm~\cite{hradil04,lvovsky04}.
The upper panel of Fig.\ref{fig:expnegat} shows that measured NPT values for such a phase-averaged tomography agree very well with theoretical expectations. It is interesting to note that, while global phase averaging lowers the NPT of the state, entanglement is still well preserved also for large mean photon numbers. Although the number of density matrix elements to reconstruct is considerably reduced with respect to the full tomography case, the largest state that we can analyze with this method has a mean photon number per mode $\bar n\approx 6$, mainly due to finite computational resources (an optimized parallel code running on a 8-CPU 3.4-GHz Xeon processor requires more than 70 hours to reconstruct a $\bar n = 6$ odd state). 

These limitations can be overcome by using a different approach.  Since the entanglement features of the odd state~(\ref{eq:oddstate}) derive from those of a displaced delocalized single photon and are therefore entirely contained in the correlated fluctuations of the quadrature measurements of the two modes, one can just use such fluctuations around their common mean values for tomographic reconstruction of the two-mode density matrix in the reduced Fock subspace of zero, one and two photons. Again, about 50 000 quadrature values are acquired for nine different relative LO phases with the global LO phase locked, and the means of the measured quadrature distributions are subtracted before reconstruction. Not requiring additional optical back displacements as in Refs.\cite{bruno13,lvovsky13,sychev18}, this scheme is free from unwanted phase and amplitude noise~\cite{macrogisin} and allows one to keep the dimensions of the reconstructed Fock space fixed regardless of the state macroscopicity, thus allowing the measurement of entanglement for very large states. 

Results are shown in the bottom panel of Fig.~\ref{fig:expnegat}. Compared to the ideal constant behavior of the theoretical NPT for the odd state shown in Fig.\ref{fig:negat}, the experimental NPT shows an unexpected decay for growing $\bar n$, but the analyzed states nonetheless preserve a relatively large degree of entanglement even for macroscopic mean photon numbers (up to $\bar n \approx 60$) in each mode. The observed degradation of the experimental NPT can be fully accounted for by including the effects of phase instabilities and limited detection and preparation efficiency. The latter has the most important effect, and we observed it decay for the state of Eq.(\ref{eq:oddstate}) while increasing the input mean photon number. The reason is a nonperfect visibility of the Mach-Zehnder herald interferometer, that makes us unable to totally erase the separable component of the state of Eq.(\ref{eq:displace}) when we set $\varphi=\pi$. This effect can be modeled by a mixing of the desired odd entangled state with a separable pair of coherent states of the same amplitude.

This approach has another interesting advantage compared to those of Refs.\cite{bruno13,lvovsky13,sychev18}. In our scheme, the two-mode entangled state is fully detected by the homodyne detector in its complete macroscopic form, and one can thus use the full quadrature measurements (comprising both the quadrature mean values and their fluctuations) in the two modes for extracting other important parameters of the state, e.g. some entanglement witness.

In conclusion, we have presented a new versatile method, based on the delocalized addition of a single photon, to entangle states of arbitrarily large size. The remarkable simplicity of our scheme will probably make it an invaluable tool for investigating the transition of quantum phenomena between the microscopic and macroscopic regimes.
By analyzing the particular case of an odd superposition of photon addition operations onto identical coherent states, we have experimentally measured significant entanglement between two modes, each populated with a mean photon number up to about 60. 
In our realization, entanglement arises from the coherent addition of a single photon over different modes. Therefore, it amounts at most to a single ebit on top of a separable state made of two relatively large coherent states. The larger the amplitude of the input coherent states, the harder it becomes to observe the quantum correlations due to the single shared photon (as we directly see from the decaying curve of the experimental NPT as a function of the coherent state size in the bottom panel of Fig.\ref{fig:expnegat}). However, in the case of $\varphi=\pi$, each mode of the entangled state contains two well-distinct components, a classical Gaussian coherent state and a nonclassical and non-Gaussian displaced single photon. These two components, although sharing a similar mean photon number (respectively, $|\alpha|^2$ or $|\alpha|^2+1$), differ significantly (and in a way that is easily distinguishable with a simple photodiode without photon-level energy resolution) in the variance of their intensity distributions (respectively proportional to $|\alpha|^2$ or $3|\alpha|^2$). According to some criteria \cite{sekatski14,macrogisin,bruno13,lvovsky13}, this is a distinctive feature of a macroscopic quantum state. Other approaches argue that different indicators (for example, the ‘‘frequency’’ of the interference fringes in the phase-space representation of the states) should be used instead to define a quantum superposition as macroscopic \cite{lee11}.
Since a definition of macroscopic quantumness and entanglement is a subtle and still quite debated issue \cite{revmacro,macrojeong}, here we only consider macroscopicity with regard to the size of the states in the two modes. This 'macro-macro' entanglement has been shown to be particularly robust against losses and is thus expected to be well suited for quantum communication tasks and storage in atomic quantum memories.

\bigskip

The authors gratefully acknowledge the support of Ente Cassa di Risparmio di Firenze under the project “MOSTO” and of the Italian Ministry of Education, University and Research (MIUR), under the “Progetto Premiale: QSecGroundSpace” and of the EU under the ERA-NET QuantERA project “ShoQC”.

 \smallskip
 \noindent  *Corresponding author. \\
  bellini@ino.it \\
 †Corresponding author.\\
 alessandro.zavatta@ino.it

\end{document}